\documentclass[conference]{IEEEtran}
\IEEEoverridecommandlockouts
\usepackage{amsmath,amssymb,amsfonts}
\usepackage{algorithmic}
\usepackage{graphicx}
\usepackage{textcomp}
\usepackage{biblatex}
\usepackage{booktabs}
\usepackage{tabularx}
\usepackage{array}

\usepackage{xcolor}
\def\BibTeX{{\rm B\kern-.05em{\sc i\kern-.025em b}\kern-.08em
    T\kern-.1667em\lower.7ex\hbox{E}\kern-.125emX}}
\begin{document}

\title{Exploiting FPGA Capabilities for Accelerated Biomedical Computing\\
\thanks{*Corresponding author}
}

\makeatletter
\newcommand{\linebreakand}{%
  \end{@IEEEauthorhalign}
  \hfill\mbox{}\par
  \mbox{}\hfill\begin{@IEEEauthorhalign}
}
\makeatother

\author{

\IEEEauthorblockN{Kayode Inadagbo*}
  \IEEEauthorblockA{\textit{Electrical and Computer Eng.} \\
    \textit{Prairie View A\&M University}\\
    Prairie View, TX \\
    kayodeinadagbo@gmail.com}
  \and
 
  \IEEEauthorblockN{Baran Arig}
  \IEEEauthorblockA{\textit{Electrical and Computer Eng.} \\
    \textit{Drexel University}\\
    Philadelphia, PA \\
    ba646@drexel.edu}
  \and
  \IEEEauthorblockN{Nisanur Alici}
  \IEEEauthorblockA{\textit{Biomedical Eng.} \\
    \textit{Erciyes University}\\
    Kayseri, Turkey \\
    alicinisanur@gmail.com}
  \linebreakand 
  \IEEEauthorblockN{Murat Isik}
  \IEEEauthorblockA{\textit{Electrical and Computer Eng.} \\
    \textit{Drexel University}\\
    Philadelphia, PA \\
    mci38@drexel.edu}
}

\IEEEoverridecommandlockouts
\maketitle
\IEEEpubidadjcol
\begin{abstract}
This study presents advanced neural network architectures including Convolutional Neural Networks (CNN), Recurrent Neural Networks (RNN), Long Short-Term Memory Networks (LSTMs), and Deep Belief Networks (DBNs) for enhanced ECG signal analysis using Field Programmable Gate Arrays (FPGAs). We utilize the MIT-BIH Arrhythmia Database for training and validation, introducing Gaussian noise to improve algorithm robustness. The implemented models feature various layers for distinct processing and classification tasks and techniques like EarlyStopping callback and Dropout layer are used to mitigate overfitting. Our work also explores the development of a custom Tensor Compute Unit (TCU) accelerator for the PYNQ Z1 board, offering comprehensive steps for FPGA-based machine learning, including setting up the Tensil toolchain in Docker, selecting architecture, configuring PS-PL, and compiling and executing models. Performance metrics such as latency and throughput are calculated for practical insights, demonstrating the potential of FPGAs in high-performance biomedical computing. The study ultimately offers a guide for optimizing neural network performance on FPGAs for various applications.

\end{abstract}


\section{Introduction}
Real-time and accurate biomedical applications, including ECG signal analysis, require high-performance computing systems. A unique combination of flexibility, performance, and energy efficiency makes Field-Programmable Gate Arrays (FPGAs) a popular choice for accelerating computations in various domains, such as biomedical engineering. In this paper, we focus on the efficient FPGA implementation of Convolutional Neural Networks (CNNs), Recurrent Neural Networks (RNNs), Long Short-Term Memory Networks (LSTMs), and Deep Belief Networks (DBNs) for ECG signal analysis tasks, including arrhythmia detection, heartbeat classification, and risk stratification. The use of FPGA-based accelerators in biomedical applications has several advantages. In addition to providing high parallelism, they are indispensable for large-scale data processing due to their ability to handle multiple tasks simultaneously. Their customizable architecture allows them to optimize performance for specific applications, which is not possible with general-purpose processors. The third advantage of FPGAs is their low latency and high bandwidth, which make them ideal for real-time processing and large data transfers. Furthermore, their energy efficiency makes them an attractive option for applications requiring significant processing power, such as image processing, machine learning, and real-time processing \cite{isik2023energy}.

In this study, we implement CNNs, RNNs, LSTMs, and DBNs in an efficient FPGA implementation using Tensil AI's open-source inference accelerator. The performance of these neural network architectures on FPGAs is optimized through various hardware designs, memory optimization techniques, and advanced compiler strategies. Our experiments demonstrate that the proposed FPGA-based accelerators maintain high accuracy with reduced precision, enabling efficient ECG signal analysis for various biomedical applications. It will become increasingly important to have high-performance computing systems that are capable of handling large amounts of data quickly and accurately as the demand for biomedical applications grows. FPGA-based accelerators can meet this demand with ease, which paves the way for further advancements in the field of biomedical engineering with our efficient implementation of CNNs, RNNs, LSTMs, and DBNs for ECG signal analysis.

The rest of the paper is organized as follows: \textbf{Section II} presents the motivation behind our work and a review of related studies in the field. \textbf{Section III} introduces open-source machine learning (ML) inference accelerators, with a focus on the Tensil AI accelerator. In \textbf{Section IV}, we describe our proposed method for efficient FPGA implementation of CNNs, RNNs, LSTMs, and DBNs for ECG signal analysis. \textbf{Section V} presents the experimental results and provides an analysis of the performance and efficiency of our proposed method. Finally, \textbf{Section VI} concludes the paper, summarizing our contributions and discussing future directions for research in this area.

\section{Motivation}
High-performance computing applications have drawn significant attention to FPGAs in recent years. An FPGA-based system can be highly optimized for specific tasks, and it can often perform better than a traditional processor. Biomedical applications of FPGAs have been increasing, providing real-time and efficient ECG signal analysis for diagnostic and monitoring purposes.

In ECG signal analysis on FPGAs, CNNs, RNNs, LSTMs, and DBNs are commonly used for detecting and classifying cardiac abnormalities. Through the analysis of ECG signals, these deep-learning techniques have shown great promise in identifying and classifying cardiac events, such as arrhythmias, myocardial infarctions, and heart failure.

A research study using CNNs on FPGAs for ECG analysis achieved high accuracy in detecting arrhythmias and classifying heartbeat types \cite{burger2020embedded}. Another study implemented an RNN on an FPGA to detect atrial fibrillation in real-time \cite{kao2020rnnaccel}. In addition to biomedical applications, FPGAs have been extensively used for high-performance computing and fault tolerance in the image and video processing industry \cite{isik2022design}. FPGAs are capable of processing high-resolution images and video in real-time. Researchers have developed an FPGA-based real-time image processing system based on Vivado HLS, a high-level synthesis tool. When filtering images, a throughput of 52 frames per second was achieved, and when segmenting images, it was 20 frames per second.

The financial industry has also used FPGAs for high-performance computing. Financial calculations often involve complex mathematical operations, which are well suited to FPGAs. With the help of FPGAs, a high-frequency trading system developed by the Tokyo Stock Exchange (TSE) can process trades in less than one microsecond. \cite{kohda2021characteristics} \cite{kohda2022characteristics}. In order to calculate financial instruments such as options and futures, FPGAs are used.

The advantages of FPGAs have made them increasingly popular in high-performance computing applications despite their challenges and limitations, such as limited on-chip memory, floating-point support, and limited availability of prebuilt IP blocks. Recent developments in high-level synthesis tools and the availability of pre-built IP blocks have made FPGAs more accessible for high-performance computing. As development tools, IP blocks, and FPGA-based solutions continue to improve, the adoption of FPGAs in high-performance computing, particularly in biomedical applications like ECG signal analysis, will increase \cite{huang2019accelerating}.

\section{Open-source ML inference accelerators}
Machine learning inference is used in many high-performance computing applications. Analyzing input data and generating output results are carried out using models. High-performance computing systems can help speed up ML inference, which is often computationally intensive. ML inference accelerators that are open-source may be beneficial to high-performance computing applications. An ML inference accelerator is a specialized hardware device that performs ML inference tasks efficiently. ML inference is typically performed on general-purpose processors or graphics processing units (GPUs) that aren't optimized for machine learning. A machine learning inference accelerator streamlines and optimizes the execution of machine learning inference tasks. A major advantage of open-source ML inference accelerators is that they are free and can be customized for specific use cases. ML inference accelerators that are open-source offer a transparent development process that encourages community participation. Furthermore, open-source accelerators can reduce the cost and time associated with the development of ML inference accelerators. An open-source ML inference accelerator, VTA (Versatile Tensor Accelerator), has recently gained significant attention. An optimized hardware accelerator is used for performing inference tasks using VTA. VTA supports TensorFlow, PyTorch, and ONNX among other ML frameworks. The VTA can be used on a variety of hardware platforms, including FPGAs and ASICs \cite{moreau2019hardware}.

\begin{figure}[h!]
    \centering
    \includegraphics[width=0.40\textwidth]{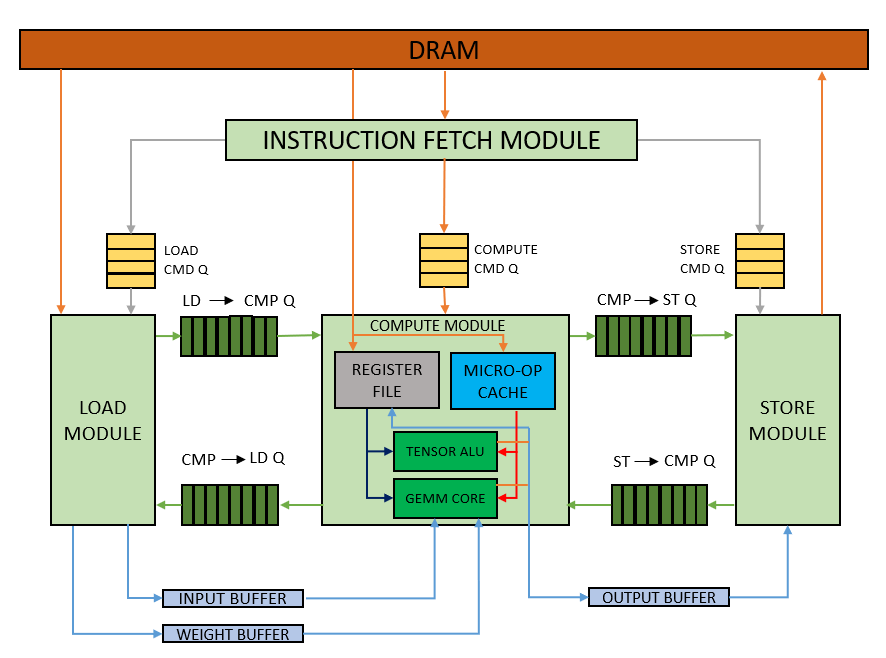}
    \caption{VTA Framework}\cite{moreau2019hardware}
    \label{fig_7}
\end{figure}

The availability of open-source tools and platforms allows developers to collaborate to develop better ML inference solutions. In addition to VTA \cite{zunin2021intel} \cite{WinNT2}, other open-source ML inference accelerators are available, such as Intel's OpenVINO and Xilinx's Deep Learning Processor. For the development and optimization of machine learning systems, these accelerators provide a variety of options. Open-source ML inference accelerators are flexible and powerful tools that can be used in high-performance computing applications. The development of ML inference systems can be customized and optimized, reducing development costs and time, fostering collaboration and innovation, and resulting in the wide adoption of ML techniques. Open-source ML inference accelerators will continue to become more powerful and efficient as the field of ML grows and evolves. The Nengo and Tensil AI frameworks can be used to build high-performance computing applications using FPGAs. Nengo software can be used to build large-scale neural models on a variety of hardware platforms, including FPGAs. Furthermore, Nengo's flexibility and extensibility let users create customized algorithms and models. Because Nengo is capable of building complex models with many neurons and synapses, it is well-suited for applications such as robotics and cognitive modeling.  Machine learning inference tasks are performed by the Tensil AI hardware accelerator. Tensil AI offers high performance at low power consumption, making it ideal for applications such as image recognition and natural language processing. Tensil AI supports a variety of machine learning frameworks, including TensorFlow and PyTorch, and can easily be integrated with existing hardware architectures. Its focus is one of the major differences between Tensil AI and Nengo. The goal of Tensil AI is to accelerate machine learning inference, whereas the goal of Nengo is to build large-scale neural networks \cite{morcos2019nengofpga} \cite{dewolf2020nengo} \cite{gosmann2017automatic} \cite{isik2023design}.  

\vspace{3pt}The versatile Nengo can be used for a wider range of tasks, while the specific Tensil AI can only be used for a limited number of tasks. The development processes of Nengo and Tensil AI are also fundamentally different. Nengo is an open-source project actively developed and maintained by a large developer community. A variety of resources, such as documentation, tutorials, and support forums, are therefore available to users. Tensil AI, on the other hand, is a commercial product that is developed and supported by the company. Users have access to dedicated support and resources, but not as much community support as with open-source software. Inference tasks in machine learning can be performed quickly and efficiently with Tensil AI. Using Tensil AI, for instance, self-driving cars and industrial automation can make inferences rapidly.  Nengo, on the other hand, simulates complex behaviors over long periods of time using large-scale neural models.  There is a potential drawback to Tensil AI, which is its limited flexibility. The specialized nature of its hardware accelerator may explain its limited versatility. Models and algorithms may not be able to be customized by users, and they may have to use prebuilt models and architectures instead. Both Nengo and Tensil AI are powerful frameworks for developing high-performance computing applications. Tensil AI is better suited for specific tasks, such as machine learning inference, than Nengo, which can be used for a wide variety of applications. It is important that developers evaluate the strengths and weaknesses of each framework carefully before making a choice, and ultimately their choice will depend on their specific application needs \cite{WinNT} \cite{WinNT1}.

\section{Method}

\subsection{Dataset}

MIT-BIH Arrhythmia Database provides valuable data on cardiac arrhythmias through the collection of ECG recordings. Between 1975 and 1979, 48 half-hour excerpts of two-channel ambulatory ECG recordings were obtained from 47 subjects at Boston's Beth Israel Hospital. These recordings were digitized at a rate of 360 samples per second per channel, with 11-bit resolution over a 10 mV range. A total of 80\% of the dataset is devoted to training, while 20\% of the dataset is devoted to validation. There are approximately 110,000 annotated beats in the original training data, which were annotated by at least two cardiologists.

The MIT-BIH Arrhythmia Database, unlike the MNIST digit dataset, does not contain noisy variations. However, if needed for a particular purpose, one can apply noise to the ECG recordings. It is possible, for instance, to simulate real-world noisy conditions in ECG data by adding random Gaussian noise to the dataset elements. In the same way that white noise follows a Gaussian distribution, Gaussian noise follows a Gaussian distribution. Impulses with random values occur.

Using the modified dataset, which now contains Gaussian noise, one can train deep-learning models for ECG signal analysis, detect arrhythmias, and improve the robustness of algorithms in noisy environments. Our dataset is shown in Fig. \ref{Figure 2}.
\begin{figure}[h]
    \graphicspath{ {D:\Stack} }
    \center \includegraphics[width=8cm, height=4cm]{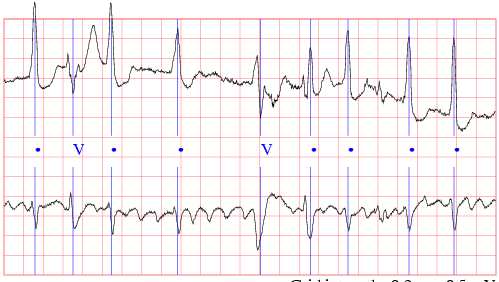}
    \caption{Project dataset}
    \label{Figure 2}
\end{figure}

\subsection{Training Details}

The code employs a four-layer neural network with LSTM for ECG signal classification. The LSTM layers understand long-term data dependencies while dense layers assist in classification. A dropout layer prevents overfitting by disabling some weights randomly, and an EarlyStopping callback halts training when validation loss stops changing. There are 64 and 32 LSTM cells in the first and second LSTM layers, respectively. The dense layer performs classification, giving 5 outputs for 5 classes. During compilation, a loss function and the 'man' optimizer are utilized for training. 'Accuracy' measures performance, and an EarlyStopping callback prevents overfitting. Training results are visualized with validation data, and the model's efficiency is evaluated using latency and throughput metrics.

The code comprises a neural network featuring several layers including a BernoulliRBM for high-level abstractions, LogisticRegression for classification, and Dropout to prevent overfitting. The Flatten layer converts 3D input to 1D, while the Reshape layer adjusts dimensions and the StandardScaler ensures data consistency. After data checks, the class imbalance is tackled with SMOTE. Deep Belief Networks (DBNs) are used on training data. The DBN model is evaluated using different sets. A CNN model classifies heartbeats in this code. Following data preprocessing and renaming of columns, the target variable distribution is plotted. Class imbalances are identified and balanced via SMOTE. The CNN model consists of Conv1D, MaxPool1D, and Dropout layers. Output is passed to a dense layer after flattening. The code offers a good start for heartbeat classification with CNN but can be improved with comments, dataset details, preprocessing steps, and hyperparameter tuning. The final part employs an RNN with a Bidirectional LSTM for ECG signal classification. The MIT-BIH Arrhythmia dataset is processed and split into training and validation sets. Model performance is assessed using latency, throughput, and other metrics.

For ECG analysis, the ideal model is one that has a low latency and a high throughput. These models can be compared based on their characteristics as follows:

\begin{itemize}

 \item LSTM (Long Short-Term Memory): ECG analysis can be performed using LSTMs when sequential data have long-term patterns, a type of RNN designed to handle long-term dependencies. Due to their complex structure, LSTMs can have higher latency and lower throughput than CNNs.

\item CNN (Convolutional Neural Network): Due to their ability to capture local patterns in data, CNNs are well suited for image and signal processing tasks. Real-time applications can benefit from their lower latency and higher throughput compared to LSTMs and RNNs.

\item RNN (Recurrent Neural Network): It is possible to model temporal dependencies in ECG signals using RNNs, which are suitable for sequential data processing. Their performance and training can be affected, however, by vanishing and exploding gradient problems. The latency and throughput of RNNs are generally higher than those of CNNs.

\item DBN (Deep Belief Network): DBNs are deep neural networks with multiple layers of Restricted Boltzmann Machines (RBMs) or autoencoders. Feature extraction and classification can be performed with them, but they aren't designed specifically for sequential data like ECG signals. DBNs may not perform as well as LSTMs or CNNs for ECG analysis, and their latency and throughput performance can vary depending on their architecture.
\end{itemize}
Based on the above comparison, CNNs are generally considered the best choice for ECG analysis when low latency and high throughput are important which is shown in Table \ref{tab:comparison}. However, the specific choice of model depends on the requirements of our application and the nature of the ECG data we are working with. It is often useful to experiment with different models to determine which one performs best for our specific use case.

\begin{table}[h]
\centering
\caption{Resource utilization summary}
\label{table:utilization}
\begin{tabular}{|c|c|c|c|}
\hline
\multicolumn{4}{|c|}{Zynq®-7000 SoC} \\ \hline
\textbf{Resource} & \textbf{Utilization} & \textbf{Available} & \textbf{\% Utilization} \\ \hline
LUT & 17579 & 74000 & 23.75 \\
FF & 20060 & 106400 & 18.85 \\
BRAM & 1374& 3300 & 41.64 \\
IO & 36 & 150 & 24 \\
DSP & 85 & 160 & 53.13 \\ \hline
\end{tabular}
\end{table}

\begin{table*}
\centering
\caption{Comparison of LSTM, CNN, RNN, and DBN models based on various parameters}
\begin{tabular}{@{}lcccccccc@{}}
\toprule
\textbf{Models} & \textbf{Accuracy} & \textbf{Precision} & \textbf{Recall} & \textbf{F1-score} & \textbf{Training time} & \textbf{Model complexity (params)} & \textbf{Throughput [GOP/s]} & \textbf{Latency} \\ \midrule
LSTM & 81\% & 28\% & 18\% & 16\% & 202.43 s & 29,477 & 2439.44 & 37 ms \\
CNN & 99\% & 20\% & 4\% & 8\% & 622.63 s & 3,245,637 & 2039.21 & 14 ms \\
RNN & 69\% & 18\% & 20\% & 19\% & 7.20 s & 276,737 & 378.10 & 43 ms \\
DBN & 19.43\% & 39\% & 20\% & 66\% & 304.18 s & 89,345 & 2237.40 & 30 ms \\ \bottomrule
\end{tabular}
\label{tab:comparison}
\end{table*}

\subsection{FPGA Implementation with Tensil's Open Source Inference Accelerator}

\begin{figure*}
    \graphicspath{ {D:\Stack} }
    \center \includegraphics[width=0.6\textwidth]{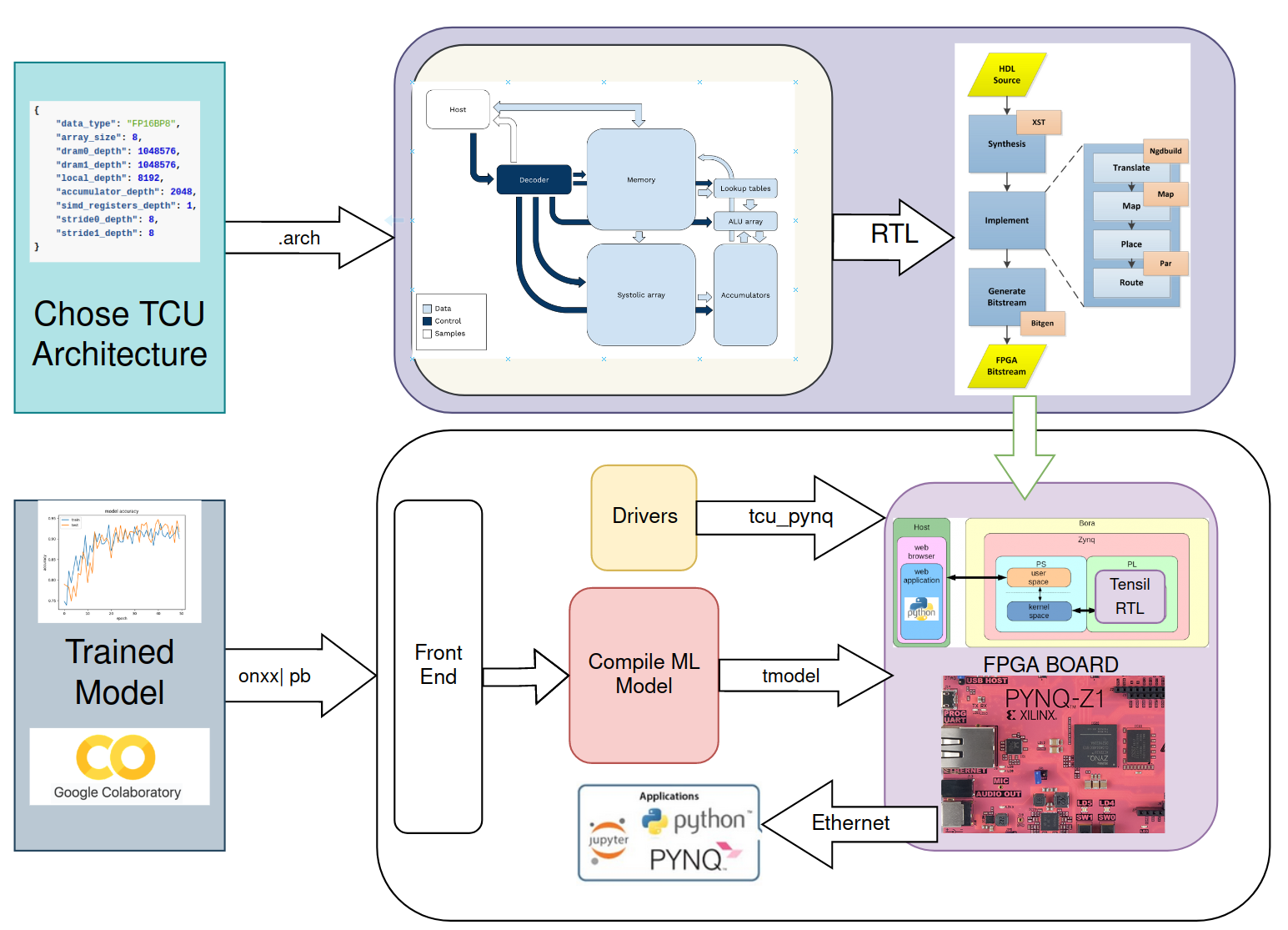}
    \caption{Block Diagram of Implementation}
    \label{Figure 3}
\end{figure*}

This section presents a detailed guide on implementing a ResNet-20 convolution model, trained not on CIFAR but the MIT-BIH Arrhythmia Database, on a PYNQ Z1 FPGA using Tensil's open-source inference accelerator, as depicted in Fig. \ref{Figure 3}. Table \ref{table:utilization} provided indicates the resource utilization on a Zynq®-7000 SoC when running the implemented model. It uses 23.75\% of available Lookup Tables (LUT), 18.85\% of Flip-Flops (FF), and 41.64\% of Block RAM (BRAM). The model also employs 24\% of available Input/Output (IO) resources and 53.13\% of Digital Signal Processors (DSP). This summary highlights the efficient use of the Zynq®-7000 SoC's resources by the model, ensuring it operates effectively within the constraints of the system.

\subsubsection{Installation and Setup of the Tensil Toolchain}

The Tensil toolchain is a suite of tools designed to facilitate FPGA development, specifically for running machine learning models on FPGA. Docker is used to host the Tensil toolchain, allowing for simple installation and setup. For those unfamiliar with Docker, it's a platform that enables developers to package and distribute their applications in a manner that is platform-independent. It's essential to ensure Docker is installed before proceeding with the Tensil toolchain setup.

\subsubsection{Selection of Architecture}

The choice of architecture depends on the specific requirements of your machine learning model and the resources available on your FPGA. It's crucial to consider factors such as the complexity of your model, the amount of data you'll be processing, and the computational resources of your FPGA.

\subsubsection{TCU Accelerator Design and Synthesis for PYNQ Z1}

In this section, a custom TCU (Tensor Compute Unit) accelerator design is generated and synthesized for the PYNQ Z1 board using the Tensil toolchain. The goal here is to create a hardware design that can efficiently execute the machine learning model on the FPGA.

\subsubsection{PS-PL Configuration and Smart Interconnect}

The next steps involve setting up the configuration between the Processing System (PS) and Programmable Logic (PL) and establishing a smart interconnect. These steps ensure efficient data transfer and proper interfacing between the various components of the design.

\subsubsection{Compilation of the ML Model}

To execute the machine learning model on the FPGA, we compile the model into a ".tmodel" file. The ".tmodel" file contains the model's structure and parameters in a format that can be executed on the FPGA. Additional files ".tprog" and ".tdata" are also generated, providing instructions and data for the FPGA execution.

\subsubsection{Execution Using PYNQ}

In this stage, all the previously generated and compiled files are deployed onto the PYNQ environment for execution. A PYNQ environment refers to a Python-based ecosystem that simplifies the usage and programming of Xilinx Zynq SoCs. It's important to set up this environment correctly on the FPGA for successful execution. An FPGA image is essentially a binary file that contains the configuration data for the FPGA. This image is crucial for defining the functionality of the FPGA hardware during runtime.

\section{Results}
Our results demonstrate the effectiveness of Tensil AI's open-source inference accelerator for optimizing neural networks and implementing them on FPGAs for high-performance computing applications. It has been done using CPUs, GPUs, and FPGAs. This research adds to the growing body of work exploring the development of FPGA-based hardware systems that support NN inference. These systems offer the promise of high throughput and power efficiency, making them a current focal point in the research community. The comparison provided in Table \ref{table:table_3} offers a comprehensive view of our approach in relation to previous implementations. The present work successfully leverages 2-D convolution, implemented on an FPGA Pynq-Z1 platform, with a ReLu activation function. This setup processes a total of 187 input samples. With an impressive number of Multiply-Accumulates (MACs) at 47,560, our approach operates at a clock speed of 100 MHz, delivering an accuracy of 99.1\% and consuming a total power of 1.53 W. These figures demonstrate the feasibility and effectiveness of our approach.

\begin{table*}
    \renewcommand{\arraystretch}{1.2}
    \setlength{\tabcolsep}{3pt}
    \centering
    \caption{Comparisons with previous implementations.}
    \label{table:table_3}
    \resizebox{0.8\linewidth}{!}{
    \begin{tabular}{|c|c|c|c|c|c|c|}
    \hline
     & \cite{zhao201913} & \cite{xia2019novel} & \cite{wang2019energy} & \cite{wong2022energy} & \cite{degirmenci2022arrhythmic} & Our \\
    \hline
    Convolution Type & 1-D & 1-D & 1-D & 2-D & 2-D & 2-D \\
    \hline
    Platform & FPGA Pynq-Z2 & CPU-i7 & - & iCE40UP5k & GPU RTX 2080 Ti& FPGA Pynq-Z1 \\
    \hline
    No. Input Samples & 512 & 200 & 400 & 10x10 & 64x64 & 187 \\
    \hline
    Activation & - & ReLu & - & bTanH & ReLu & ReLu  \\
    \hline
    Num of MACs & 929,650 & 1,289,312 & 749,620 & 27,153 & 58.1 M & 47,560
  \\
    \hline
    Clock & 25 MHz & 3.7 GHz & - & 100 MHz & 1350 MHz & 100 MHz  \\
    \hline
    Accuracy & 98.9 & 99.8 & 98.4 & 96.8 & 99.7 & 99.1  \\
    \hline
    Power & 13.34 $\mu$W & 84 W & 141 mW & 227.3 $\mu$W & 108 W & 1.53 W  \\
    \hline
    \end{tabular}}
\end{table*}

\section{Conclusions}

The extensive guide for running ResNet-20 convolution models on a PYNQ Z1 FPGA with Tensil's open-source inference accelerator exhibits the potential for FPGAs to speed up machine learning tasks. The exploration demonstrated how the Tensil toolchain, along with specific hardware architecture selection, meticulous model compilation, and effective FPGA execution setup, can deliver outstanding performance and accuracy. The project also demonstrated the inherent advantages of FPGAs, including low latency and optimized memory usage, which led to impressive results. An energy-efficient, reconfigurable system was established by successfully implementing the machine learning model across heterogeneous devices. Our next step is to incorporate Dynamic Partial Reconfiguration (DPR), an advanced technology in the field of reconfigurable hardware, into the system framework in order to further improve performance. We will use Tensil AI to reshape and offload initial and post-data processing stages in order to facilitate high-performance computing. Tensil AI offers a flexible platform that allows the manipulation of inputs and outputs to accommodate different models by compiling new ones, which has proven to be advantageous. We anticipate further advances in machine learning implementations on FPGAs as we continue to develop our methods and technologies.

\printbibliography

\end{document}